\documentclass[aps,prb,reprint,showpacs,nofootinbib]{revtex4-1}

\usepackage{graphicx}
\usepackage{amsmath}
\usepackage{amsfonts}
\usepackage[overlay,absolute]{textpos}

\begin{document}

\begin{textblock*}{\textwidth}[0,0](19mm,7mm)
\center\footnotesize\noindent
To appear in Phys.\ Rev.\ B: https://journals.aps.org/prb/
\end{textblock*}

\title{Anyon braiding in semi-analytical fractional quantum Hall lattice models}

\author{Anne E. B. Nielsen}
\affiliation{Max-Planck-Institut f\"ur Quantenoptik, Hans-Kopfermann-Str.\ 1, D-85748 Garching, Germany}

\begin{abstract}
It has been demonstrated numerically, mainly by considering ground state properties, that fractional quantum Hall physics can appear in lattice systems, but it is very difficult to study the anyons directly. Here, I propose to solve this problem by using conformal field theory to build semi-analytical fractional quantum Hall lattice models having anyons in their ground states, and I carry out the construction explicitly for the family of bosonic and fermionic Laughlin states. This enables me to show directly that the braiding properties of the anyons are those expected from analytical continuation of the wave functions and to compute properties such as internal structure, size, and charge of the anyons with simple Monte Carlo simulations. The models can also be used to study how the anyons behave when they approach or even pass through the edge of the sample. Finally, I compute the effective magnetic field seen by the anyons, which varies periodically due to the presence of the lattice.
\end{abstract}


\maketitle

The discovery of the fractional quantum Hall (FQH) effect \cite{tsui} revealed the existence of phases of matter that are fundamentally different from previously known phases. These phases have attracted much attention both because new physics is needed to describe them \cite{wen} and because their properties are interesting for quantum computing \cite{nayak}. With the aim of getting a deeper understanding of the effect and find more robust and controllable ways to realize it experimentally, much effort is currently being put into exploring under which conditions the effect occurs. A major result in this direction is the discovery that FQH physics can be realized in lattice systems,\cite{sorensen,hafezi,kapit,neupert,sheng,wangyf,regnault,nielsen12,%
tu13,nielsen13,cincio,tu14,bauer} which opens up doors towards investigating the effect under new parameter regimes, maybe even room temperature \cite{tang}.

One of the special features of FQH states is the possibility to create anyons. Anyons are particle-like excitations that have a more complicated exchange statistics than bosons and fermions. By now, more techniques have been developed that allow one to determine which types of anyons can be created in a system by looking only at the properties of the ground states, \cite{niu,wen90,hamma1,hamma2,hatsugai,kitaev,levin,li,wangh,zhang,cincio,jiang,TuMP} and these techniques have been used to demonstrate the FQH nature of the above mentioned lattice models. The techniques do, however, have limitations in that they do not provide information about, e.g., what the size and internal structure of the anyons are, how the anyons can be created and moved around, and the details of braiding operations. In order to describe these features, one needs to study the anyons directly, but this is very difficult for lattice FQH models, where only the Hamiltonian (or at best the Hamiltonian and the anyon free ground state \cite{schroeter,thomale,kapit,nielsen12,tu13,tu14}) is known analytically (although test computations can be done for very small systems \cite{kapitb}).

In this article, I provide a solution to these problems by proposing to construct lattice FQH models that have anyons in their ground states and for which both the Hamiltonian and the ground states are expressed analytically. This allows me to study the anyons in great detail, even for quite large systems, by using simple Monte Carlo simulations. In addition, it is immediately clear how to move the anyons around, since the positions of the anyons are parameters of the Hamiltonian.

Analytical wave functions have played an important role in understanding the FQH effect in continuum systems,\cite{laughlin,haldane83,halperin,jain,moore} and it is also possible to write down states, whose analytical continuation properties suggest that the states contain anyons. It has, however, taken a long time and a lot of effort to confirm numerically that braiding anyons in these states in fact leads to the same changes of the wave functions as doing analytical continuation alone.\cite{arovas,gurarie,paredes,read,bonderson,bernevig} The states proposed here can be seen as lattice versions of the continuum states, but because the states are now defined on lattices, it is numerically easy to do braiding operations and check that the changes of the states are as expected from analytical continuation.

Another unusual feature of the models proposed here is that the Hamiltonians are still exact if the coordinates of the anyons are taken out of the sample. This gives excellent possibilities for studying how the properties of the anyons change, when the anyons approach or pass through the edge of the sample. It also suggests an alternative viewpoint, in which anyons are not excitations in a system, but instead normal particles that become anyons when moving on a background of other particles. This idea may be interesting to explore in proposals for realizing anyons experimentally.

The construction I propose builds on the idea of expressing wave functions in terms of conformal field theory (CFT) correlators \cite{cirac,moore} and the idea of using CFT properties to derive parent Hamiltonians for such states in the continuum\cite{moore} or the lattice\cite{nielsen11}. Here, I demonstrate that CFT properties can also be used to derive parent Hamiltonians of states containing anyons. I specifically construct models, whose ground states are lattice Laughlin states with filling factor $1/q$, $q\in\mathbb{N}$, containing quasi-holes, but it is likely that the same approach can be used to build several other lattice FQH models with Abelian and non-Abelian anyons. It is, e.g., already known how to construct a number of different FQH states in the continuum in terms of CFT correlators, and these states can be easily transformed to lattice states.

\textit{Wave functions.--}%
I first use CFT to construct lattice Laughlin states containing quasi-holes. Consider an arbitrary lattice in two dimensions with lattice sites at the positions $z_j$, $j=1,2,\ldots,N$, where $z_j$ are complex numbers. The positions of the quasi-holes are likewise specified by $w_j$, $j=1,2,\ldots,Q$. To each of the lattice sites I associate a vertex operator $V_{n_j}(z_j)=(-1)^{(j-1)n_j}{:\exp[i(qn_j-1)\phi(z_j)/\sqrt{q}]:}$, and to each of the quasi-holes I associate the vertex operator $W_{p_j}(w_j)={:\exp[ip_j\phi(w_j)/\sqrt{q}]:}$. Here, $\phi(z)$ is the chiral part of the field of a free massless boson, $:\ldots:$ stands for normal ordering, $q$ is a positive integer, $n_j\in\{0,1\}$ is the number of hard-core bosons/fermions at lattice site number $j$ for $q$ even/odd, and, as we shall see below, $p_j/q$ with $p_j\in\{1,2,\ldots,q-1\}$ is the charge of the quasi-hole at $w_j$. The wave function is then defined as
\begin{equation}\label{wf}
|\Psi_q\rangle=\sum_{n_1,\ldots,n_N}\Psi_q(w_{1\to Q},n_{1\to N})|n_1,\ldots,n_N\rangle,
\end{equation}
where $w_{1\to Q}$ is shorthand for $w_1,w_2,\ldots,w_Q$,
\begin{multline}\label{CFTcor}
\Psi_q(w_{1\to Q},n_{1\to N})\propto
\langle0| W_{p_1}(w_1)W_{p_2}(w_2)\cdots W_{p_Q}(w_Q)\\
\times V_{n_1}(z_1)V_{n_2}(z_2)\cdots V_{n_N}(z_N)|0\rangle,
\end{multline}
and $\langle0|\ldots|0\rangle$ denotes the vacuum expectation value in the CFT. Note that the positions $z_i$ of the lattice sites are fixed throughout, whereas the positions $w_i$ of the quasi-holes are taken to be parameters. A quasi-hole coordinate $w_i$ may coincide with one of the lattice sites. In that case the model is the same as the model obtained by leaving out the lattice site and placing a quasi-hole with charge $(p_i-1)/q$ at the position.

Evaluating \eqref{CFTcor} using standard methods\cite{CFTbook} gives
\begin{multline}\label{Laughlin}
\Psi_q(w_{1\to Q},n_{1\to N})=\mathcal{C}(w_{1\to Q})^{-1}\delta_{n}
\prod_{i<j}(w_i-w_j)^{p_ip_j/q}\\
\times\prod_{i,j}(w_i-z_j)^{p_in_j}
\prod_{i<j}(z_{i}-z_{j})^{qn_{i}n_{j}}\\
\times\prod_{i,j}(w_i-z_j)^{-p_i/q}
\prod_{i\neq j}(z_{i}-z_{j})^{-n_{i}},
\end{multline}
where $\mathcal{C}$ is a real normalization constant and $\delta_n=1$ for $\sum_{j=1}^Nn_j=(N-\sum_{j=1}^Qp_j)/q$ and $\delta_n=0$ otherwise. Note that it is not required that $\sum_{j=1}^Qp_j$ is itself divisible by $q$. It is therefore possible to have, e.g., models with just one quasi-hole with charge $1/q$.

The construction above is reminiscent of the corresponding construction for continuum Laughlin states with quasi-holes proposed in \onlinecite{moore}, but there are important differences in the way charge neutrality is ensured. Note also that if the lattice is defined on a disc shaped region and the area per lattice site $a$ is the same for all sites, then the norm of $\prod_{i,j}(w_i-z_j)^{-p_i/q}\prod_{i\neq j}(z_{i}-z_{j})^{-n_{i}}$ approaches the usual Gaussian factor $\exp(-\frac{1}{4}\frac{2\pi}{a}\sum_{i=1}^Q\frac{p_i}{q}|w_i|^2 -\frac{1}{4}\frac{2\pi}{a}\sum_{i=1}^N|z_i|^2)$ in the thermodynamic limit (and the phase can be transformed away if desired).\cite{nielsen12,tu14}

\begin{figure}
\includegraphics[width=0.9\columnwidth,bb=186 283 438 534,clip]{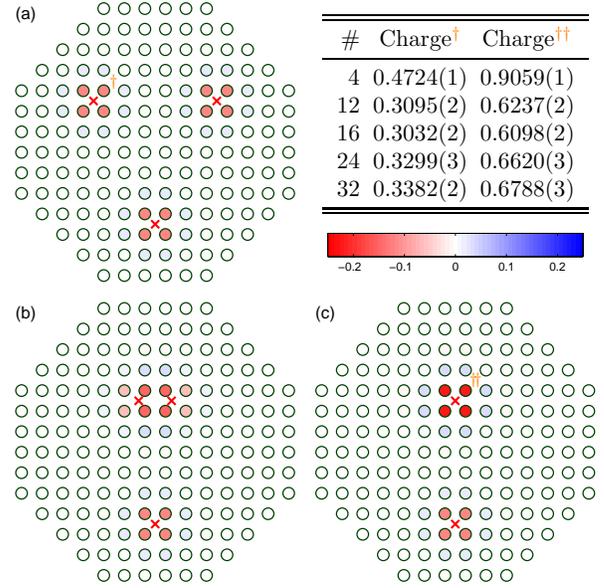}
\caption{(a) Difference $\Delta n_j\equiv \langle n_j\rangle_{Q=3}-\langle n_j\rangle_{Q=0}$ between the particle density of the state with three quasi-holes with $p_i=1$ and the state without quasi-holes for $q=3$. The red crosses mark the positions of the quasi-holes, each ring is a lattice site, and the color is the density difference. (b-c) Illustration of how the density difference changes when two quasi-holes are brought together to fuse to a single quasi-hole. The table shows minus the sum of $\Delta n_j$ over the $\#$ lattice sites closest to the quasi-hole marked with $\dag$ ($\dag\dag$). This quantity approaches the charge of the quasi-hole for $\#$ large. (The numbers in parentheses are the uncertainties on the last digit estimated from the Monte Carlo simulations.)}\label{fig:density}
\end{figure}

\textit{Parent Hamiltonians.--}%
The next step is to derive a Hamiltonian, for which \eqref{wf} is the exact ground state. Our starting point is the fact that the field
\begin{equation}\label{null}
\chi(z_i)=\oint_{z_i}\frac{dz}{2\pi i}\frac{1}{z-z_i}
[G^{+}(z)V_{-}(z_i)-qJ(z)V_{+}(z_i)]
\end{equation}
is a null field.\cite{tu14} Here, $G^+(z)={:e^{i\sqrt{q}\phi(z)}:}$, $J(z)=i\partial_z \phi(z)/\sqrt{q}$, $V_-(z)={:e^{-i\phi(z)/\sqrt{q}}:}$, and
$V_+(z)={:e^{i(q-1)\phi(z)/\sqrt{q}}:}$. Therefore
\begin{multline}\label{dec}
\langle W_{p_1}(w_1)\cdots W_{p_Q}(w_Q)
V_{n_1}(z_1)\cdots V_{n_{i-1}}(z_{i-1}) \chi(z_i)\\
\times V_{n_{i+1}}(z_{i+1}) \cdots V_{n_N}(z_N)\rangle=0.
\end{multline}
Using standard tools from CFT and complex analysis (the technical details can be found in the Supplemental Material below), \eqref{dec} can be rewritten into $\Lambda_i|\Psi\rangle=0$, where
\begin{equation}\label{Lambda}
\Lambda_i=\sum_{j(\neq i)}\frac{1}{z_i-z_j}[d_i^\dag d_j-n_i(qn_j-1)]
-\sum_j\frac{p_j}{z_i-w_j}n_i.
\end{equation}
$d_j$ is the hard-core boson/fermion annihilation operator acting on site $j$ for $q$ even/odd and $n_j=d_j^\dag d_j$. In addition, $\left[\sum_{i=1}^N n_i-(N-\sum_{j=1}^Qp_j)/q\right]|\Psi\rangle=0$ due to the $\delta_n$ factor in \eqref{Laughlin}. The positive semi-definite operator
\begin{equation}\label{Ham}
H=\sum_i\Lambda_i^\dag\Lambda_i+c [\sum_{i=1}^N n_i-(N-\sum_{j=1}^Qp_j)/q]^2,
\end{equation}
where $c$ is a positive constant, is therefore a parent Hamiltonian of \eqref{wf}. Note that $c\rightarrow\infty$ corresponds to fixing the number of particles in the system. I have confirmed numerically for a number of small lattices with $q=3$ and $q=4$ that the ground state is unique.

The Hamiltonian \eqref{Ham} contains only one-, two-, and three-body terms. It is quite common that Hamiltonians with FQH ground states contain three-body or higher interactions, and this is one of the motivations for the significant current efforts towards finding suitable ways to realize three-body interactions in optical lattices.\cite{buchler,mazza,daley} The Hamiltonian is also seen to involve interactions between distant sites in the system. It has, however, been found for related models that the Hamiltonian can be transformed into a local Hamiltonian without significantly altering the ground state,\cite{nielsen13,tusun} and this suggests that there is a chance that the same is true here. One of the local Hamiltonians has, in addition, been used to propose an implementation scheme for a FQH lattice model in ultracold atoms in optical lattices.\cite{nielsen13,nielsen14}

\begin{figure}
\includegraphics[width=0.9\columnwidth,bb=118 269 472 568,clip]{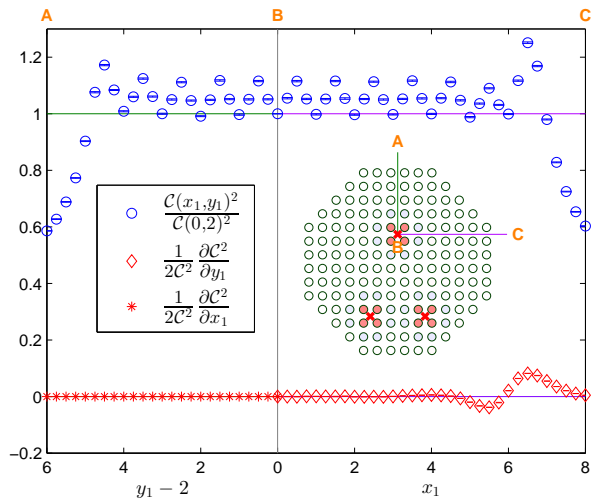}
\caption{Normalization constant $\mathcal{C}^2$ and the derivatives $(2\mathcal{C}^2)^{-1} \partial \mathcal{C}^2/\partial y_1$ and $(2\mathcal{C}^2)^{-1} \partial \mathcal{C}^2/\partial x_1$ appearing in \eqref{theta2} for the state with $q=3$ and $3$ quasi-holes with charge $1/3$ when the quasi-hole at $w_1=x_1+iy_1$ is moved along the path from $A$ to $B$ to $C$ as shown in the inset. The $x_1$ derivative along the path from $A$ to $B$ is zero due to symmetry.}\label{fig:inout}
\end{figure}

\textit{Quasi-holes.--}%
In the following, I use the Metropolis Monte Carlo algorithm to investigate important properties of the quasi-holes in the models. I shall consider $q=3$ and the square lattice in Fig.~\ref{fig:density} with $N=156$ throughout, and since the state \eqref{Laughlin} is invariant under scale transformations, I shall arbitrarily set the lattice constant to one. I start with the internal structure and charge of the quasi-holes. Figure~\ref{fig:density} shows the difference between the particle density of the state with three quasi-holes and the particle density of the state without quasi-holes for different choices of $w_{1\to3}$. It is seen that the quasi-holes are screened and have a diameter of a few lattice constants. The figure also illustrates how the difference in density changes when two of the quasi-holes are brought close together and fused to a single quasi-hole.

If the fermionic particles are imagined to have charge $-1$ as in the FQH effect, then the density difference is minus the excess charge, and the charges of the quasi-holes can be determined by adding up the excess charges in a region around $w_i$. Quasi-holes with $p_i=1$ ($p_i=2$) are expected to have a charge of $1/3$ ($2/3$), and this is consistent with the results obtained in the figure. It is also interesting to note that the charge distribution of the quasi-hole with $p_i=2$ is not just two times the charge distribution of the quasi-hole with $p_i=1$.

\begin{figure}
\includegraphics[width=0.9\columnwidth,bb=103 268 471 571,clip]{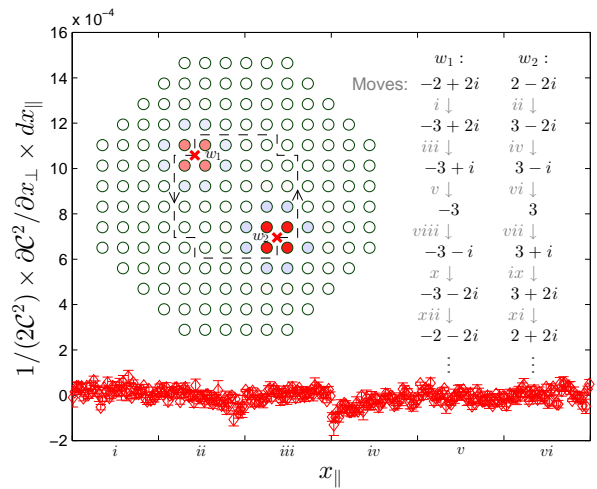}
\caption{Braiding of two quasi-holes with charges $1/3$ and $2/3$ in the model with $q=3$. The quasi-holes are moved along the path shown with a dashed line. First $w_1$ is moved one lattice spacing to the left with $w_2$ fixed, then $w_2$ is moved one lattice spacing to the right with $w_1$ fixed and so on following the moves listed on the right (the dots stand for another 36 moves following the same pattern as the first 12 moves but with the figure rotated $90^\circ$, $180^\circ$, and $270^\circ$, respectively). The plot shows the contributions to the integral in \eqref{theta2} for the first 6 moves ($x_\parallel$ is the coordinate along the curve, $x_\perp$ is the coordinate perpendicular to the curve in the inward direction, and I have chosen $|dx_\parallel|=0.02$). Due to symmetry, the integral over the complete path is 8 times the integral over the first 6 moves. Adding up the contributions and multiplying by 8, I get $\theta=-0.0017(5)\times 2\pi$, where the error is the statistical error of the Monte Carlo computations.}\label{fig:braid}
\end{figure}

\begin{figure*}
\includegraphics[width=0.9\textwidth,bb=48 511 539 666,clip]{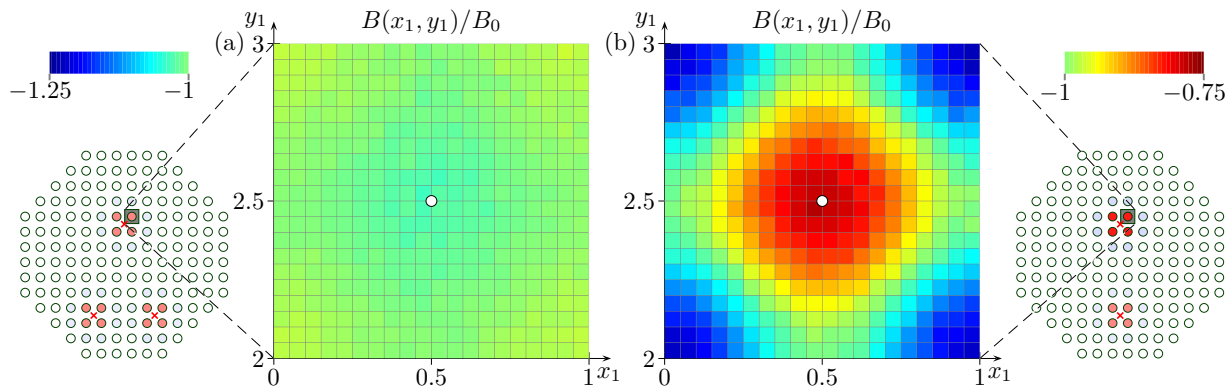}
\caption{The effective magnetic field seen by a quasi-hole at the position $w_1=x_1+iy_1$ for $q=3$ when the charge of the quasi-hole is $1/3$ (a) or $2/3$ (b) and the quasi-hole is well-separated from the other quasi-holes in the system. $w_1$ is chosen to be within one unit cell as shown, and the integral of $B$ over the unit cell is $-1.000\,B_0$ for both plots.}\label{fig:field}
\end{figure*}

\textit{Braiding.--}%
When braiding quasi-holes, one adiabatically moves the coordinates $w_{1\to Q}$ along some closed path. This transforms the wave function as $|\Psi_q\rangle\rightarrow Me^{i\theta}|\Psi_q\rangle$, where $Me^{i\theta}$ is the Berry phase factor \cite{berry,read},
\begin{equation}\label{theta1}
\theta=i\sum_{j=1}^Q\oint_c \Big(\langle\Psi_q|\frac{\partial \Psi_q}{\partial w_j}\rangle dw_j
+\langle\Psi_q|\frac{\partial \Psi_q}{\partial \bar{w}_j}\rangle d\bar{w}_j\Big),
\end{equation}
and $M$ is the monodromy, i.e.\ the change obtained from analytical continuation alone.

The monodromy can be determined by inspection of \eqref{Laughlin}. When the $i$th quasi-hole is moved in the counter-clockwise direction around the $j$th quasi-hole, the wave function picks up the factor $e^{2\pi i p_ip_j/q}$, and when the $i$th quasi-hole is moved in the counter-clockwise direction around a lattice site, the wave function picks up the factor $e^{-2\pi i p_i /q}$. The former is the expected braiding statistics of the quasi-holes, and the latter is the Aharonov-Bohm phase of a particle with charge $p_ie/q$ moving around a loop enclosing a magnetic flux of $-h/e$, where $h$ is Planck's constant and $e$ is the elementary charge.

For the state \eqref{Laughlin}, \eqref{theta1} simplifies to
\begin{equation}\label{theta2}
\theta=\frac{1}{2}\sum_{j=1}^Q \oint_c \left(
\frac{1}{\mathcal{C}^2}\frac{\partial \mathcal{C}^2}{\partial y_j}dx_j
-\frac{1}{\mathcal{C}^2}\frac{\partial \mathcal{C}^2}{\partial x_j}dy_j
\right),
\end{equation}
where $w_j=x_j+iy_j$. Computing the derivatives analytically (using \eqref{Laughlin}), one can express the integrant in terms of $\langle n_k\rangle$, which is easily evaluated with Monte Carlo. Monte Carlo can also be used to compute $\mathcal{C}^2$ up to a constant factor that does not depend on $w_{1\to Q}$.

The results provided in Fig.~\ref{fig:inout} show that the integrant in \eqref{theta2} is practically zero and $\mathcal{C}^2$ varies with the period of the lattice as long as the quasi-hole that is being moved stays well inside the sample, is sufficiently far from the other quasi-holes, and moves along lines midway between the lattice sites. Under these conditions, the Berry phase factor hence equals the monodromy.

I next consider braiding explicitly as shown in Fig.~\ref{fig:braid}. Here, a quasi-hole with charge $1/3$ moves around $32$ lattice sites, a quasi-hole with charge $2/3$ moves around $32$ lattice sites, and a quasi-hole with charge $1/3$ moves around a quasi-hole with charge $2/3$. For this process, I also get that $\theta$ is practically zero (up to expected finite size effects), and the Berry phase factor is therefore practically the monodromy, i.e.\ $\exp[2\pi i(-32+2/3)]$.

\textit{Effective magnetic field.--}%
It follows from the results above that the magnetic flux through one unit cell is $-h/e$. As opposed to the continuum case, however, the magnetic field does not need to be uniform. To get a more detailed picture, consider a quasi-hole that moves around a loop that does not enclose other quasi-holes. From the divergence theorem, $\theta=-(1/2)\iint \nabla^2\ln[\mathcal{C}(x_1,y_1)^2] dx_1dy_1$, where the integral is over the area enclosed by the loop. On the other hand, if the Berry phase is interpreted as an Aharonov-Bohm phase of a charged particle in a magnetic field $B$, then $\theta-i\ln(M)=(p_1e/q)(2\pi/h)a\iint B(x_1,y_1)dx_1dy_1$. Therefore, after using \eqref{Laughlin} to rewrite $\nabla^2\ln(\mathcal{C}^2)$,
\begin{equation}
\frac{B}{B_0}=-\frac{qp_1}{\pi}\sum_{j,k}
\frac{\langle n_jn_k\rangle-\langle n_j\rangle \langle n_k\rangle} {(w_1-z_k)(w_1-z_j)^*},
\end{equation}
where $B_0=h/(ea)$. Results for $p_1=1$ and $p_1=2$ are shown in Fig.~\ref{fig:field}. Despite the presence of the lattice, it is observed that the magnetic field is not too far from uniform with variations of up to $4\%$ for the quasi-hole with $p_1=1$ and up to $23\%$ for the quasi-hole with $p_1=2$. It is also interesting that although the fields are the same on average, the local fields seen by the two types of quasi-holes differ, which is a result of the different ways in which the quasi-holes affect their environment.

\textit{Conclusion.--}%
Due to the complexity of many-body systems, analytical models are particularly helpful to gain insight. Here, I have constructed a model with an analytical ground state and Hamiltonian, which makes it possible to study lattice Laughlin anyons in great detail with simple numerical computations. In future work, I plan to extend the above construction to build different fractional quantum Hall lattice models with Abelian and non-Abelian anyons.

\textit{Acknowledgment.--}%
The author would like to thank J.\ Ignacio Cirac and Germ\'an Sierra for discussions on related topics.

\bibliography{bibfil}

\newpage

\onecolumngrid
\newpage

\begin{center}
\textbf{Supplemental Material}
\end{center}

\section*{Operator annihilating the wave function with $Q$ quasi-holes}

From Eqs.~\eqref{null} and \eqref{dec} in the main article, it follows that
\begin{multline}\label{start}
\oint_{z_i}\frac{dz}{2\pi i}\frac{1}{z-z_i}
\langle W_{p_1}(w_1)\ldots W_{p_Q}(w_Q)
V_{n_{1}}(z_{1})\ldots G^{+}(z)V_{-}(z_i) \ldots V_{n_{N}}(z_{N})\rangle\\
-q\oint_{z_i}\frac{dz}{2\pi i}\frac{1}{z-z_i}\langle
W_{p_1}(w_1)\ldots W_{p_Q}(w_Q)
V_{n_{1}}(z_{1})\ldots J(z)V_+(z_i) \ldots V_{n_{N}}(z_{N})\rangle=0.
\end{multline}
The first term on the left hand side of \eqref{start} can be rewritten using the operator product expansions \cite{CFTbook}
\begin{equation}
G^{+}(z)W_{p_j}(w_j) \sim 0, \qquad
G^{+}(z)V_{n_j}(z_j) \sim (-1)^{(j-1)}\frac{\sum_{n'_j}g_{n_jn'_j}}{z-z_j}V_{n'_j}(z_j),
\end{equation}
where $g_{01}=1$ and $g_{00}=g_{10}=g_{11}=0$. From $:e^{i\alpha\phi(z)}::e^{i\beta\phi(w)}:{}=
(z-w)^{\alpha\beta}:e^{i\alpha\phi(z)+i\beta\phi(w)}:$, it follows that $:e^{i\alpha\phi(z)}::e^{i\beta\phi(w)}:{}=
(-1)^{\alpha\beta}:e^{i\beta\phi(w)}::e^{i\alpha\phi(z)}:$, and in particular
\begin{equation}
V_{n_j}(z_j)G^{+}(z)=(-1)^{qn_j-1}G^{+}(z)V_{n_j}(z_j).
\end{equation}
Therefore
\begin{align}
&\oint_{z_i}\frac{dz}{2\pi i}\frac{1}{z-z_i}
\langle W_{p_1}(w_1)\ldots W_{p_Q}(w_Q)
V_{n_{1}}(z_{1})\ldots G^{+}(z)V_{-}(z_i) \ldots V_{n_{N}}(z_{N})\rangle\nonumber\\
&=-\sum_{j(\neq i)}\oint_{z_j}\frac{dz}{2\pi i}\frac{1}{z-z_i}
\langle W_{p_1}(w_1)\ldots W_{p_Q}(w_Q)
V_{n_{1}}(z_{1})\ldots G^{+}(z)V_{-}(z_i) \ldots V_{n_{N}}(z_{N})\rangle\nonumber\\
&=-(-1)^{i-1}\sum_{j=1}^{i-1}\oint_{z_j}\frac{dz}{2\pi i}\frac{(-1)^{q\sum_{k=j}^{i-1}n_k}}{z-z_i}
\frac{\sum_{n'_j}g_{n_jn'_j}}{z-z_j}
\langle W_{p_1}(w_1)\ldots W_{p_Q}(w_Q)
V_{n_{1}}(z_{1})\ldots V_{n'_j}(z_j) \ldots V_{-}(z_i) \ldots V_{n_{N}}(z_{N})\rangle\nonumber\\
&\phantom{=}-(-1)^{i-1}\sum_{j=i+1}^N\oint_{z_j}\frac{dz}{2\pi i}\frac{(-1)^{q\sum_{k=i+1}^{j-1}n_k}}{z-z_i}
\frac{\sum_{n'_j}g_{n_jn'_j}}{z-z_j}
\langle W_{p_1}(w_1)\ldots W_{p_Q}(w_Q)
V_{n_{1}}(z_{1})\ldots V_{-}(z_i) \ldots V_{n'_j}(z_j) \ldots V_{n_{N}}(z_{N})\rangle\nonumber\\
&=-(-1)^{i-1}\sum_{j=1}^{i-1}\frac{(-1)^{q\sum_{k=j}^{i-1}n_k}}{z_j-z_i}
\sum_{n'_j}g_{n_jn'_j}
\langle W_{p_1}(w_1)\ldots W_{p_Q}(w_Q)
V_{n_{1}}(z_{1})\ldots V_{n'_j}(z_j) \ldots V_{-}(z_i) \ldots V_{n_{N}}(z_{N})\rangle\nonumber\\
&\phantom{=}-(-1)^{i-1}\sum_{j=i+1}^N\frac{(-1)^{q\sum_{k=i+1}^{j-1}n_k}}{z_j-z_i}
\sum_{n'_j}g_{n_jn'_j}
\langle W_{p_1}(w_1)\ldots W_{p_Q}(w_Q)
V_{n_{1}}(z_{1})\ldots V_{-}(z_i) \ldots V_{n'_j}(z_j) \ldots V_{n_{N}}(z_{N})\rangle.
\end{align}
Multiplying this expression by $(-1)^{i-1}|n_1,\ldots,n_{i-1},1,n_{i+1}\ldots,n_N\rangle =(-1)^{i-1}\sum_{n'_i,n_i}g_{n'_in_i}|n_1,\ldots,n_i,\ldots,n_N\rangle$ and summing over all $n_k$, $k\neq i$, gives
\begin{equation}\label{firstterm}
\sum_{j(\neq i)}\frac{d_i^\dag d_j}{z_i-z_j}|\Psi\rangle,
\end{equation}
where $d_j$ is the hard-core boson/fermion annihilation operator acting on site $j$ for $q$ even/odd.

The second term on the left hand side of \eqref{start} can be rewritten using the operator product expansion \cite{CFTbook}
\begin{equation}
J(z) :e^{i\alpha\phi(z_j)}:{} \sim \frac{1}{\sqrt{q}}
\frac{\alpha}{z-z_j}:e^{i\alpha\phi(z_j)}:.
\end{equation}
Specifically,
\begin{align}
&-q\oint_{z_i}\frac{dz}{2\pi i}\frac{1}{z-z_i}\langle
W_{p_1}(w_1)\ldots W_{p_Q}(w_Q)
V_{n_{1}}(z_{1})\ldots J(z)V_+(z_i) \ldots V_{n_{N}}(z_{N})\rangle\nonumber\\
&=q\sum_{j(\neq i)}\oint_{z_j}\frac{dz}{2\pi i}\frac{1}{z-z_i}\langle
W_{p_1}(w_1)\ldots W_{p_Q}(w_Q)
V_{n_{1}}(z_{1})\ldots J(z)V_+(z_i) \ldots V_{n_{N}}(z_{N})\rangle\nonumber\\
&\phantom{=}+q\sum_{j}\oint_{w_j}\frac{dz}{2\pi i}\frac{1}{z-z_i}\langle
W_{p_1}(w_1)\ldots W_{p_Q}(w_Q)
V_{n_{1}}(z_{1})\ldots J(z)V_+(z_i) \ldots V_{n_{N}}(z_{N})\rangle\nonumber\\
&=\sum_{j(\neq i)}\frac{qn_j-1}{z_j-z_i}\langle
W_{p_1}(w_1)\ldots W_{p_Q}(w_Q)
V_{n_{1}}(z_{1}) \ldots V_+(z_i) \ldots V_{n_{N}}(z_{N})\rangle\nonumber\\
&\phantom{=}+\sum_{j}\frac{p_j}{w_j-z_i}\langle
W_{p_1}(w_1)\ldots W_{p_Q}(w_Q)
V_{n_{1}}(z_{1})\ldots V_+(z_i) \ldots V_{n_{N}}(z_{N})\rangle\nonumber.
\end{align}
Multiplying this expression by $(-1)^{i-1}|n_1,\ldots,n_{i-1},1,n_{i+1}\ldots,n_N\rangle =(-1)^{i-1}\sum_{n_i}n_i|n_1,\ldots,n_i,\ldots,n_N\rangle$ and summing over all $n_k$, $k\neq i$, gives
\begin{equation}\label{secondterm}
-\sum_{j(\neq i)}\frac{qn_j-1}{z_i-z_j}n_i|\Psi\rangle -\sum_{j}\frac{p_j}{z_i-w_j}n_i|\Psi\rangle.
\end{equation}
Since the sum of \eqref{firstterm} and \eqref{secondterm} is zero, it follows that $\Lambda_i|\Psi\rangle=0$ with $\Lambda_i$ given in Eq.~\eqref{Lambda} in the main text.

\end{document}